\documentclass[conference]{IEEEtran}
\usepackage{amsmath}
\usepackage{latexsym}
\usepackage{graphicx}
\usepackage{bbding}
\usepackage{indentfirst}
\IEEEoverridecommandlockouts
\begin{document}

\title{On the Secrecy Outage Capacity of Physical Layer Security in Large-Scale MIMO Relaying Systems with Imperfect CSI}
\author{\authorblockN{Xiaoming~Chen$^{\dagger,\ddagger}$, Lei~Lei$^{\dagger}$, Huazi~Zhang$^{\star}$ and Chau~Yuen$^{\ast}$
\\$^{\dagger}$ College of Electronic and Information Engineering, Nanjing
University of Aeronautics and Astronautics, China.
\\$^{\ddagger}$ National Mobile Communications Research Laboratory, Southeast
University, China.
\\$^{\star}$ Department of ECE, North Carolina State University, USA.
\\$^{\ast}$ Singapore University of Technology and Design, Singapore.
\\Email: \{chenxiaoming,leilei\}@nuaa.edu.cn, hzhang17@ncsu.edu, yuenchau@sutd.edu.sg
\thanks{This work was supported by the National Natural
Science Foundation of China (No. 61301102, 61100195), the Natural
Science Foundation of Jiangsu Province (No. BK20130820), the open
research fund of National Mobile Communications Research Laboratory,
Southeast University (No. 2012D16), the Doctoral Fund of Ministry of
Education of China (No. 20123218120022) and the iTrust.}}}
\maketitle

\begin{abstract}
In this paper, we study the problem of physical layer security in a
large-scale multiple-input multiple-output (LS-MIMO) relaying
system. The advantage of LS-MIMO relaying systems is exploited to
enhance both wireless security and spectral efficiency. In
particular, the challenging issue incurred by short interception
distance is well addressed. Under very practical assumptions, i.e.,
no eavesdropper's channel state information (CSI) and imperfect
legitimate channel CSI, this paper gives a thorough investigation of
the impact of imperfect CSI in two classic relaying systems, i.e.,
amplify-and-forward (AF) and decode-and-forward (DF) systems, and
obtain explicit expressions of secrecy outage capacities for both
cases. Finally, our theoretical claims are validated by the
numerical results.
\end{abstract}

\section{Introduction}
The open nature of wireless channel gives rise to many wireless
security problems. Traditionally, wireless security is guaranteed
through high layer encryption. With the development of interception
technology, encryption becomes more complex, leading to high
computation burden. Thanks to the physical layer security measures
enlightened by information theory, we may exploit wireless channel
characteristics to guarantee secure communications \cite{Wyner} even
without encryption.

The performance of physical layer security is measured by secrecy
rate, namely the capacity difference between the legitimate channel
from the information source to the destination and the eavesdropper
channel from the information source to the eavesdropper \cite{SC1}
\cite{SC2}. As expected, the introduction of relay into physical
layer security can improve the legitimate channel capacity through
cooperative diversity, and thus enhances transmission security
\cite{Relay}. Some feasible relaying schemes and their performances
were discussed in \cite{RelayingSchemes}. Relay positions were then
optimized from the the perspective of minimizing the interception
probability in \cite{Placement}. Further, best relay selection was
proposed to suppress the interception probability by exploiting
selective gain in multiple relay systems \cite{RelaySelection}.
Additionally, multiple relay cooperative beamforming combined with
jamming was also adopted to maximize the secrecy rate
\cite{Multi-Relay}. In fact, if the relay is equipped with multiple
antennas, more effective relaying strategy can be used to optimize
the secrecy rate \cite{MIMORelay1}. Linear precoding schemes were
investigated in a MIMO relay network assuming global CSI
\cite{MIMORelay2}. Nevertheless, due to the fact that the
eavesdroppers are usually well hidden, it is difficult for the relay
to obtain the knowledge of the eavesdropper channel. Therefore, it
is difficult to realize absolutely secure communications over the
fading channel. Under this condition, the conception of secrecy
outage capacity was adopted to guarantee a secure communication with
a high probability \cite{SOC1} \cite{SOC2}.

A challenging problem rises when the eavesdropper is close to the
information transmitter. Even with a multi-antenna relay, the
secrecy outage capacity is too small to fulfill the requirement of
quality of service (QoS), due to the relatively high quality of
intercepted signal. Recently, it is found that large-scale MIMO
(LS-MIMO) systems can significantly improve the transmission
performance by utilizing its enormous array gain
\cite{LargescaleMIMO1} \cite{LargescaleMIMO2}. Inspired by this, we
propose the use of an LS-MIMO relay to enhance wireless security.
Note that a critical issue of the LS-MIMO system is that channel
state information (CSI) may be imperfect due to duplex delay in time
division duplex (TDD) systems, resulting in inevitable performance
loss. In this paper, we focus on the performance analysis in terms
of secrecy outage capacity in an LS-MIMO relay system with imperfect
CSI under the amplify-to-forward (AF) and the decode-to-forward (DF)
relaying strategies, respectively. The contributions of this paper
are two-fold:
\begin{enumerate}
\item To the best of our knowledge, we are the first to introduce
LS-MIMO into a relay system, and solve the challenging problem of
short-distance interception.

\item We derive explicit expressions of secrecy outage capacity
for AF and DF relaying systems, and then provide some guidelines for
performance optimization.
\end{enumerate}

The rest of this paper is organized as follows. We first give an
overview of the LS-MIMO relaying system employing physical layer
security in Section II, and then derive the secrecy outage
capacities based on AF and DF relaying strategies with imperfect CSI
in Section III. In Section IV, we present some numerical results to
validate the effectiveness of the proposed scheme. Finally, we
conclude the whole paper in Section V.

\section{System Model}
\begin{figure}[h] \centering
\includegraphics [width=0.4\textwidth] {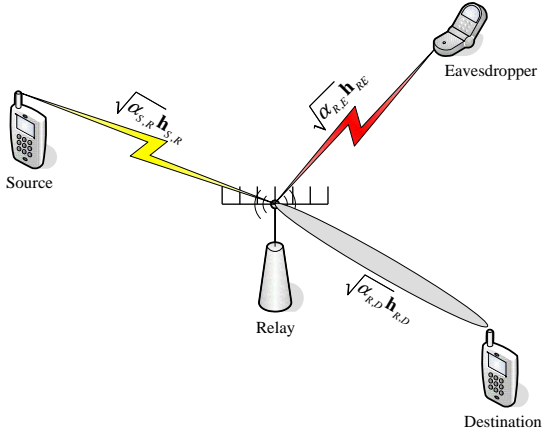}
\caption {An overview of the large-scale MIMO relaying system employing physical layer security.}
\label{Fig1}
\end{figure}

We consider a time division duplex (TDD) LS-MIMO relaying system,
including one source, one destination equipped with a single antenna
each and one relay deploying $N_R$ antennas in presence of a passive
single antenna eavesdropper, as shown in Fig.\ref{Fig1}. It is worth
pointing out that $N_R$ is usually quite large in such an LS-MIMO
relaying system, i.e. $N_R=100$ or greater. The system works in the
half-duplex mode, so the information transmission from the source to
the destination via the aid of the relay requires two time slots.
Specifically, in the first time slot, the source transmits the
signal to the relay, and then the relay forwards the post-processing
signal to the destination during the second time slot. Note that the
direct link from the source to the destination is unavailable due to
long distance. Meanwhile, the eavesdropper monitors the transmission
from the relay to the destination, and tries to intercept the
signal. Following \cite{RelaySelection}, we also assume that the
eavesdropper is out of the coverage area of the source, since it
thought the information comes from the relay.

We use $\sqrt{\alpha_{i,j}}\textbf{h}_{i,j}$ to denote the channel
from $i$ to $j$, where $i\in\{S,R\}$, $j\in\{R,D,E\}$ and $S, R, D,
E$ represent the source, the relay, the destination and the
eavesdropper, respectively. $\alpha_{i,j}$ is the distance-dependent
path loss and $\textbf{h}_{i,j}$ is the channel small scale fading.
In this paper, we model $\textbf{h}_{i,j}$ as Gaussian distribution
with zero mean and unit variance. $\alpha_{i,j}$ remains constant
during a relatively long period and $\textbf{h}_{i,j}$ fades
independently slot by slot. Thus, the receive signals at the relay
in the first time slot can be expressed as
\begin{equation}
\textbf{y}_R=\sqrt{P_S\alpha_{S,R}}\textbf{h}_{S,R}s+\textbf{n}_R,\label{eqn1}
\end{equation}
where $s$ is the normalized Gaussian distributed transmit signal,
$P_S$ is the transmit power at the source, $\textbf{n}_R$ is the
additive Gaussian white noise with zero mean and unit variance at
the relay. Through post-processing to $\textbf{y}_R$ according to
the CSI $\textbf{h}_{S,R}$ and $\textbf{h}_{R,D}$, the relay
forwards a normalized signal $\textbf{r}$ to the destination with
power $P_R$, then the received signals at the destination and the
eavesdropper are given by
\begin{equation}
y_D=\sqrt{P_R\alpha_{R,D}}\textbf{h}_{R,D}^H\textbf{r}+n_D,\label{eqn3}
\end{equation}
and
\begin{equation}
y_{E}=\sqrt{P_R\alpha_{R,E}}\textbf{h}_{R,E}^H\textbf{r}+n_{E},\label{eqn4}
\end{equation}
respectively, where $n_D$ and $n_{E}$ are the additive Gaussian
white noises with zero mean and unit variance at the destination and
the eavesdropper.

Assuming the legitimate channel and the eavesdropper channel
capacities are $C_D$ and $C_E$, from the perspective of information
theory, the secrecy capacity is given by $C_{SEC}=[C_D-C_E]^+$,
where $[x]^+=\max(x,0)$ \cite{SC1}. Since there is no knowledge of
the eavesdropper channel at the source and relay, it is impossible
to provide a steady secrecy capacity. In this paper, we take the
secrecy outage capacity $C_{SOC}$ as the performance metric, which
is defined as the maximum rate under the condition that the outage
probability that the transmission rate surpasses the secrecy
capacity is equal to a given value $\varepsilon$, namely
\begin{equation}
P_r(C_{SOC}>C_D-C_E)=\varepsilon.\label{eqn5}
\end{equation}

\section{Secrecy Outage Capacity Analysis}
In this section, we concentrate on the analysis of secrecy outage
capacity of physical layer security in an LS-MIMO relaying system.
Considering that AF and DF are two commonly used relaying
strategies, we study the two cases in sequence.

\subsection{Amplify-and-Forward (AF) Case}
In this case, the relay forwards the signal $\textbf{r}^{AF}$ via
multiplying the received signal $\textbf{y}_R$ by a $N_R\times N_R$
processing matrix $\textbf{F}$, namely
\begin{equation}
\textbf{r}^{AF}=\textbf{F}\textbf{y}_R.\label{eqn6}
\end{equation}
We assume the relay has full CSI $\textbf{h}_{S,R}$ by channel
estimation, and gets partial CSI $\textbf{h}_{R,D}$ by via channel
reciprocity in TDD systems. Due to duplex delay between uplink and
downlink, there is a certain degree of mismatch between the
estimated CSI $\hat{\textbf{h}}_{R,D}$ and the real CSI
$\textbf{h}_{R,D}$, whose relation can be expressed as
\cite{CSIMismatch}
\begin{equation}
\textbf{h}_{R,D}=\sqrt{\rho}\hat{\textbf{h}}_{R,D}+\sqrt{1-\rho}\textbf{e},\label{eqn7}
\end{equation}
where $\textbf{e}$ is the error noise vector with independent and
identically distributed (i.i.d.) zero mean and unit variance complex
Gaussian entries. $\rho$, scaling from 0 to 1, is the correlation
coefficient between $\hat{\textbf{h}}_{R,D}$ and $\textbf{h}_{R,D}$.
A large $\rho$ means better CSI accuracy. If $\rho=1$, the relay has
full CSI $\textbf{h}_{R,D}$. Additionally, due to the hidden
property of the eavesdropper, the CSI $\textbf{h}_{R,E}$ is
unavailable. Therefore, $\textbf{F}$ is designed only based on
$\textbf{h}_{S,R}$ and $\hat{\textbf{h}}_{R,D}$, but is independent
of $\textbf{h}_{R,E}$. Considering the better performance of maximum
ratio combination (MRC) and maximum ratio transmission (MRT) in
LS-MIMO systems, we design $\textbf{F}$ by combining MRC and MRT.
Specifically, the received signal $\textbf{y}_R$ is combined with an
MRC vector $\frac{\textbf{h}_{S,R}^H}{\|\textbf{h}_{S,R}\|}$, and
then normalized by a scaling factor
$\frac{1}{\sqrt{P_S\alpha_{S,R}\|\textbf{h}_{S,R}\|^2+1}}$, and
finally multiplied by an MRT vector
$\frac{\hat{\textbf{h}}_{R,D}}{\|\hat{\textbf{h}}_{R,D}\|}$. In
other words, the processing matrix is given by
\begin{equation}
\textbf{F}=\frac{\hat{\textbf{h}}_{R,D}}{\|\hat{\textbf{h}}_{R,D}\|}\frac{1}{\sqrt{P_S\alpha_{S,R}\|\textbf{h}_{S,R}\|^2+1}}\frac{\textbf{h}_{S,R}^H}{\|\textbf{h}_{S,R}\|}.\label{eqn8}
\end{equation}
Thus, the received signal at the destination and the corresponding
signal-to-noise ratio (SNR) can be expressed as
\begin{eqnarray}
y_D^{AF}&=&\frac{\sqrt{P_SP_R\alpha_{S,R}\alpha_{R,D}}\textbf{h}_{R,D}^H\hat{\textbf{h}}_{R,D}\textbf{h}_{S,R}^H\textbf{h}_{S,R}}{\|\hat{\textbf{h}}_{R,D}\|\sqrt{P_S\alpha_{S,R}\|\textbf{h}_{S,R}\|^2+1}\|\textbf{h}_{S,R}\|}s\nonumber\\
&+&\frac{\sqrt{P_R\alpha_{R,D}}\textbf{h}_{R,D}^H\hat{\textbf{h}}_{R,D}\textbf{h}_{S,R}^H}{\|\hat{\textbf{h}}_{R,D}\|\sqrt{P_S\alpha_{S,R}\|\textbf{h}_{S,R}\|^2+1}\|\textbf{h}_{S,R}\|}\textbf{n}_R+n_D,\nonumber\\\label{eqn9}
\end{eqnarray}
and
\begin{eqnarray}
\gamma_D^{AF}&=&\frac{\frac{P_SP_R\alpha_{S,R}\alpha_{R,D}|\textbf{h}_{R,D}^H\hat{\textbf{h}}_{R,D}|^2\|\textbf{h}_{S,R}\|^2}{\|\hat{\textbf{h}}_{R,D}\|^2(P_S\alpha_{S,R}\|\textbf{h}_{S,R}\|^2+1)}}{\frac{P_R\alpha_{R,D}|\textbf{h}_{R,D}^H\hat{\textbf{h}}_{R,D}|^2}{\|\hat{\textbf{h}}_{R,D}\|^2(P_S\alpha_{S,R}\|\textbf{h}_{S,R}\|^2+1)}+1}\nonumber\\
&=&\frac{a|\textbf{h}_{R,D}^H\hat{\textbf{h}}_{R,D}|^2\|\textbf{h}_{S,R}\|^2}{b|\textbf{h}_{R,D}^H\hat{\textbf{h}}_{R,D}|^2+\|\hat{\textbf{h}}_{R,D}\|^2(c\|\textbf{h}_{S,R}\|^2+1)},\label{eqn10}
\end{eqnarray}
where $a=P_SP_R\alpha_{S,R}\alpha_{R,D}$, $b=P_R\alpha_{R,D}$ and
$c=P_S\alpha_{S,R}$. Based on the receive SNR in (\ref{eqn10}), we
have the following theorem:

\emph{Theorem 1}: The legitimate channel capacity in an LS-MIMO
relaying system in presence of imperfect CSI can be approximated as
$C_D^{AF}=W\log_2\left(1+\frac{P_SP_R\alpha_{S,R}\alpha_{R,D}\rho
N_R^2}{P_R\alpha_{R,D}\rho N_R+P_S\alpha_{S,R}N_R+1}\right)$, where
$W$ is the spectral bandwidth.

\begin{proof}
Please refer to Appendix I.
\end{proof}

It is found that the legitimate channel capacity is a constant due
to channel hardening in such an LS-MIMO relaying system. Similar to
(\ref{eqn10}), the SNR at the eavesdropper is given by
\begin{equation}
\gamma_E^{AF}=\frac{d
|\textbf{h}_{R,E}^H\hat{\textbf{h}}_{R,D}|^2\|\textbf{h}_{S,R}\|^2}
{e|\textbf{h}_{R,E}^H\hat{\textbf{h}}_{R,D}|^2
+\|\hat{\textbf{h}}_{R,D}\|^2(c\|\textbf{h}_{S,R}\|^2+1)},\label{eqn11}
\end{equation}
where $d=P_SP_R\alpha_{S,R}\alpha_{R,E}$ and $e=P_R\alpha_{R,E}$.
Hence, according to the definition of secrecy outage capacity in
(\ref{eqn5}), we have the following theorem:

\emph{Theorem 2}: Given the outage probability bound by
$\varepsilon$, the secrecy outage capacity based on the AF relaying
strategy is
$C_{SOC}^{AF}=C_D^{AF}-W\log_2\left(1+\frac{dN_R\ln\varepsilon}{e\ln\varepsilon-cN_R-1}\right)$.

\begin{proof}
Please refer to Appendix II.
\end{proof}

Based on Theorem 2, we may obtain the interception probability
$P_{0}^{AF}$, namely the probability that the legitimated channel
capacity is less than the eavesdropper channel capacity. By letting
$C_{SOC}^{AF}=0$ in (\ref{app4}), we have
\begin{eqnarray}
P_{0}^{AF}&=&1-F\left(2^{C_{D}^{AF}/W}-1\right)\nonumber\\
&=&\exp\left(-\frac{(cN_R+1)\left(2^{C_{D}^{AF}/W}-1\right)}{dN_R-e\left(2^{C_{D}/W}-1\right)}\right).\label{eqn12}
\end{eqnarray}

\subsection{Decode-and-Forward (DF) Case}
Different from the AF relaying strategy, the DF decodes the receive
signal at the relay, and then forwards the original signal to the
destination, so as to avoid noise amplification. Also by using the
MRC technique at the relay, the channel capacity from the source to
the relay can be expressed as
\begin{eqnarray}
C_{S,R}^{DF}&=&W\log_2(1+P_S\alpha_{S,R}\|\textbf{h}_{S,R}\|^2)\nonumber\\
&=&W\log_2(1+P_S\alpha_{S,R}N_R),\label{eqn13}
\end{eqnarray}
where (\ref{eqn13}) holds true based on channel hardening when
$N_R\rightarrow\infty$. Then, the relay performs MRT based on the
estimated CSI $\hat{\textbf{h}}_{R,D}$. The channel capacity from
the relay to the destination is given by
\begin{eqnarray}
C_{R,D}^{DF}&=&W\log_2\left(1+P_R\alpha_{R,D}\left|\textbf{h}_{R,D}^H\frac{\hat{\textbf{h}}_{R,D}}{\|\hat{\textbf{h}}_{R,D}\|}\right|^2\right)\nonumber\\
&\approx&W\log_2\left(1+P_R\alpha_{R,D}\rho N_R\right),\label{eqn14}
\end{eqnarray}
where (\ref{eqn14}) is obtained similarly to Theorem 1. Thus, the
legitimate channel capacity under the DF relaying strategy can be
expressed as
\begin{eqnarray}
C_{D}^{DF}&=&\min(C_{S,R}^{DF},C_{R,D}^{DF})\nonumber\\
&=&W\log_2\left(1+\min(P_S\alpha_{S,R}N_R,P_R\alpha_{R,D}\rho
N_R)\right).\nonumber\\\label{eqn15}
\end{eqnarray}
It is found that $C_D^{DF}$ is also a constant due to channel
hardening in an LS-MIMO system. Meanwhile, the eavesdropper
intercepts the signal from the relay, the corresponding channel
capacity can be computed as
\begin{eqnarray}
C_{E}^{DF}=W\log_2\left(1+P_R\alpha_{R,E}\left|\textbf{h}_{R,E}^H\frac{\hat{\textbf{h}}_{R,E}}{\|\hat{\textbf{h}}_{R,D}\|}\right|^2\right).\label{eqn16}
\end{eqnarray}
For the secrecy outage capacity in an LS-MIMO DF relaying system, we
have the following theorem:

\emph{Theorem 3}: Given the outage probability bound by
$\varepsilon$, the secrecy outage capacity based on the DF relaying
strategy is
$C_{SOC}^{DF}=C_D^{DF}-W\log_2\left(1+P_R\alpha_{R,E}\ln\varepsilon\right)$.

\begin{proof}
Please refer to Appendix III.
\end{proof}

Similarly, we can also obtain the interception probability in this
case. Let $C_{SOC}^{DF}=0$ in (\ref{app9}), we have
\begin{eqnarray}
P_0^{DF}=\exp\left(-\frac{2^{C_D^{DF}/W}-1}{P_R\alpha_{R,E}}\right).\label{eqn17}
\end{eqnarray}

\section{Numerical Results}
To examine the accuracy of the derived theoretical expressions of
the secrecy outage capacity for the LS-MIMO AF and DF relaying
systems, we present several numerical results in the following
scenarios: we set $N_R=100$, $W=10$KHz and $\rho=0.9$. The relay is
in the middle of a line between the source and the destination. We
normalize the path loss as $\alpha_{S,R}=\alpha_{R,D}=1$ and use
$\alpha_{R,E}$ to denote the relative path loss. For example, if
$\alpha_{R,E}>1$, then the eavesdropper is closer to the relay than
the destination. In addition, we use SNR$_S=10\log_{10}P_S$ and
SNR$_R=10\log_{10}P_R$ to represent the transmit signal-to-noise
ratio (SNR) in dB at the source and the relay, respectively.

\begin{figure}[h] \centering
\includegraphics [width=0.5\textwidth] {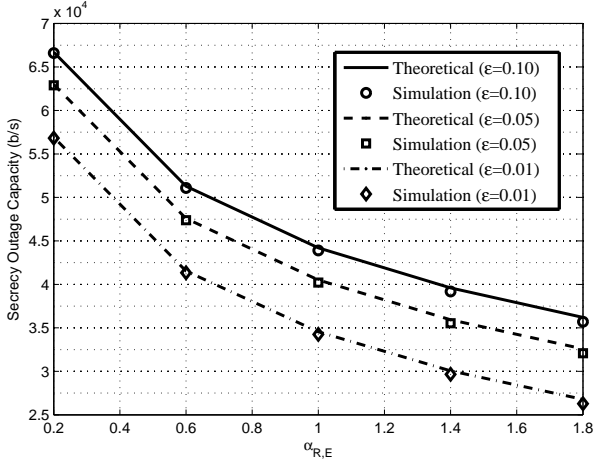}
\caption {Comparison of theoretical and simulation results based on
AF relaying mode.} \label{Fig2}
\end{figure}

Firstly, we testify the accuracy of the theoretical expression in AF
relaying mode with SNR$_s=$SNR$_R=20$dB. As seen in Fig.\ref{Fig2},
the theoretical results are well consistent with the simulations in
the whole $\alpha_{R,E}$ region with different outage probability
requirements, which proves the high accuracy of the derived
performance expressions. Given the outage probability bound by
$\varepsilon$, as $\alpha_{R,E}$ increases, the secrecy outage
capacity decreases gradually, this is because the interception
ability of the eavesdropper enhances due to the short interception
distance. In addition, given $\alpha_{R,E}$, the secrecy outage
capacity improves with the increase of $\varepsilon$, since the
outage probability is an increasing function of the secrecy outage
capacity.

\begin{figure}[h] \centering
\includegraphics [width=0.5\textwidth] {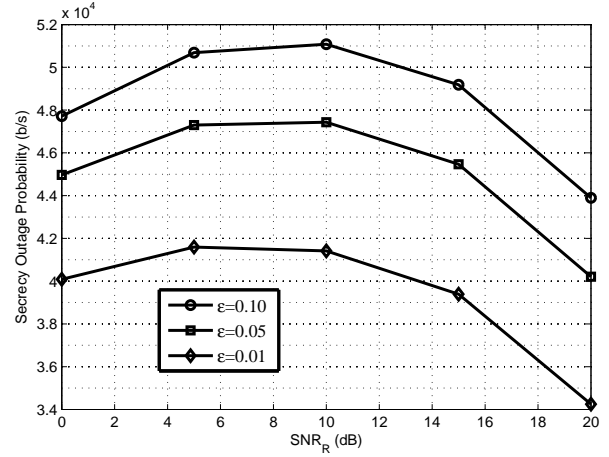}
\caption {Performance comparison based on AF relaying mode with
different SNR$_R$.} \label{Fig6}
\end{figure}

Secondly, we show the impact of SNR$_R$ on the secrecy outage
capacity in AF relaying mode with $\varepsilon=0.01$,
$\alpha_{R,E}=1$, and SNR$_S=20$dB. As seen in Fig.\ref{Fig6}, the
secrecy outage capacity is not an increasing function of SNR$_R$,
since both the legitimate and eavesdropper channel capacities
improve as SNR$_R$. Thus, it makes sense to find the optimal SNR$_R$
to maximize the secrecy outage capacity in LS-MIMO relaying systems.

\begin{figure}[h] \centering
\includegraphics [width=0.5\textwidth] {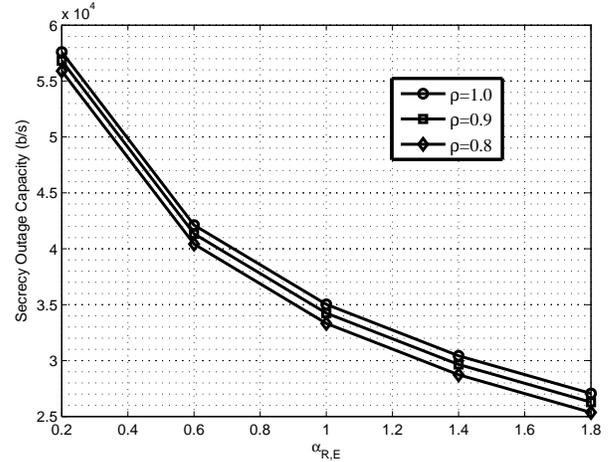}
\caption {Performance comparison based on AF relaying mode with
different $\rho$.} \label{Fig4}
\end{figure}

Thirdly, we investigate the impact of $\rho$ on the secrecy outage
capacity in AF relaying mode with SNR$_s=$SNR$_R=20$dB and
$\varepsilon=0.01$. The CSI mismatch will result in the performance
loss. However, in LS-MIMO relay systems, when the number of antennas
is quite large, the impact of CSI mismatch can be weakened due to
high spatial resolution. As shown in Fig.\ref{Fig4}, the performance
loss by reducing $\rho$ from 1 to 0.8 is slight. In other words, the
AF relaying scheme is insensitive to the CSI accuracy.

\begin{figure}[h] \centering
\includegraphics [width=0.5\textwidth] {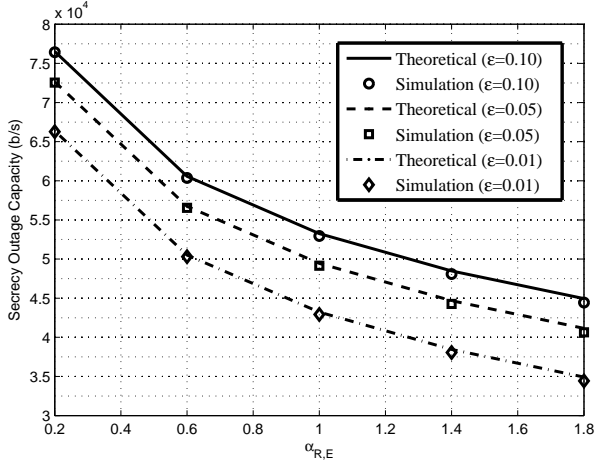}
\caption {Comparison of theoretical and simulation results based on
DF relaying mode.} \label{Fig3}
\end{figure}

Then, we testify the accuracy of the derived theoretical expressions
based on DF relaying mode with SNR$_s=$SNR$_R=20$dB. As seen in
Fig.\ref{Fig3}, the theoretical results coincide with the
simulations nicely. Similar to the AF relaying mode, the secrecy
outage capacity decreases as $\alpha_{R,E}$ increases and
$\varepsilon$ reduces. Note that, compared to the secrecy outage
capacity of AF relaying mode in Fig.\ref{Fig2}, the secrecy outage
capacity of DF relaying mode is better under the same conditions,
since the DF mode avoids amplifying the noise at the relay.

\begin{figure}[h] \centering
\includegraphics [width=0.5\textwidth] {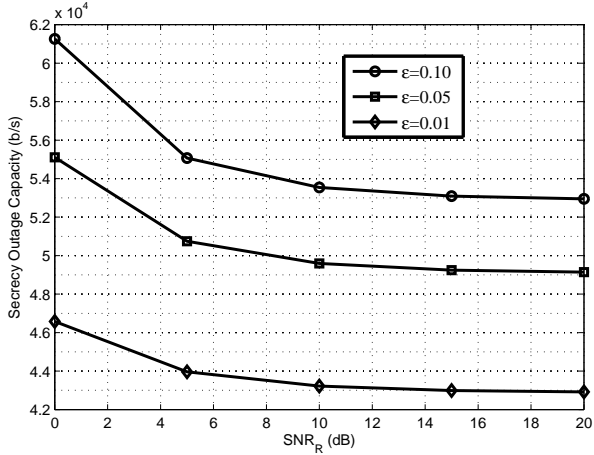}
\caption {Performance comparison based on DF relaying mode with
different SNR$_R$.} \label{Fig7}
\end{figure}

Next, we investigate the impact of SNR$_R$ of the secrecy outage
capacity in DF relaying mode with $\varepsilon=0.01$,
$\alpha_{R,E}=1$, and SNR$_S=20$dB. From Fig.\ref{Fig7}, as SNR$_R$
increases, unexpectedly the secrecy outage capacity decreases. This
is because the legitimate channel capacity is independent of SNR$_R$
when SNR$_R$ is large, but the eavesdropper channel capacity is an
increasing function of SNR$_R$. Hence, it is also necessary to
choose an optimal SNR$_R$ to optimize the performance.

\begin{figure}[h] \centering
\includegraphics [width=0.5\textwidth] {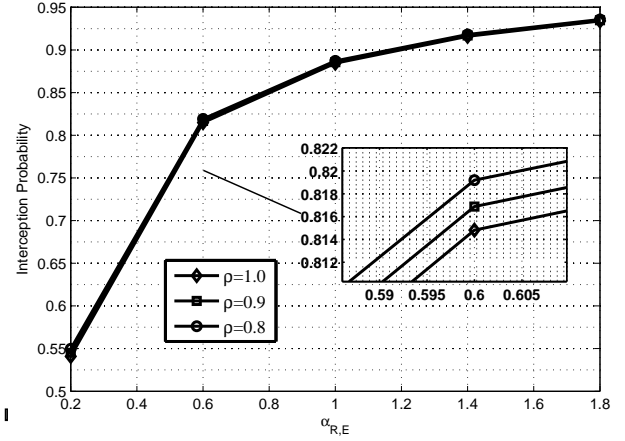}
\caption {Interception probability comparison based on DF relaying
mode with different $\rho$.} \label{Fig5}
\end{figure}

Finally, we show the impact of $\rho$ on the interception
probability based on DF relaying mode with SNR$_s=$SNR$_R=20$dB and
$\varepsilon=0.01$. As seen in Fig.\ref{Fig5}, although the
interception probability increases as $\rho$ decreases, the
performance gap is nearly negligible, so the DF relaying scheme has
high robustness.

\section{Conclusion}
A major contribution of this paper is the introduction of the
LS-MIMO relaying technique into physical layer security to
significantly enhance wireless security, especially when the
interception distance is short. This paper focuses on the analysis
of secrecy outage capacity under the AF and DF relaying strategies,
and derives the closed-form expressions in terms of the transmit SNR
and the channel condition. Altogether, these results provide some
important guidelines for performance optimization of physical layer
security in LS-MIMO relaying systems.

\begin{appendices}
\section{Proof of Theorem 1}
Based on the SNR $\gamma_D^{AF}$ at the destination, the legitimated
channel capacity can be expressed as
\begin{eqnarray}
C_D^{AF}&=&W\log_2(1+a|\textbf{h}_{R,D}^H\hat{\textbf{h}}_{R,D}|^2\|\textbf{h}_{S,R}\|^2\nonumber\\
&/&(b|\textbf{h}_{R,D}^H\hat{\textbf{h}}_{R,D}|^2+\|\hat{\textbf{h}}_{R,D}\|^2(c\|\textbf{h}_{S,R}\|^2+1)))\nonumber\\
&=&W\log_2\bigg(1+a\left|(\sqrt{\rho}\hat{\textbf{h}}_{R,D}+\sqrt{1-\rho}\textbf{e})^H\frac{\hat{\textbf{h}}_{R,D}}{\|\hat{\textbf{h}}_{R,D}\|}\right|^2\nonumber\\
&*&\|\textbf{h}_{S,R}\|^2/\bigg(b\left|(\sqrt{\rho}\hat{\textbf{h}}_{R,D}+\sqrt{1-\rho}\textbf{e})^H\frac{\hat{\textbf{h}}_{R,D}}{\|\hat{\textbf{h}}_{R,D}\|}\right|^2\nonumber\\
&+&(c\|\textbf{h}_{S,R}\|^2+1)\bigg)\bigg)\label{app1}\\
&=&W\log_2\bigg(1+a(\rho\|\hat{\textbf{h}}_{R,D}\|^2+2\sqrt{\rho(1-\rho)}\mathcal{R}(\textbf{e}^H\hat{\textbf{h}}_{R,D})\nonumber\\
&+&(1-\rho)\|\textbf{e}\hat{\textbf{h}}_{R,D}^H\|^2/\|\hat{\textbf{h}}_{R,D}\|^2)\|\textbf{h}_{S,R}\|^2/(b(\rho\|\hat{\textbf{h}}_{R,D}\|^2\nonumber\\
&+&2\sqrt{\rho(1-\rho)}\mathcal{R}(\textbf{e}^H\hat{\textbf{h}}_{R,D})+(1-\rho)\|\textbf{e}\hat{\textbf{h}}_{R,D}^H\|^2/\|\hat{\textbf{h}}_{R,D}\|^2)\nonumber\\
&+&(c\|\textbf{h}_{S,R}\|^2+1)\bigg)\bigg)\nonumber\\
&\approx&W\log_2\left(1+\frac{a\rho\|\hat{\textbf{h}}_{R,D}\|^2\|\textbf{h}_{S,R}\|^2}{b\rho\|\hat{\textbf{h}}_{R,D}\|^2+c\|\textbf{h}_{S,R}\|^2+1}\right)\label{app2}\\
&\approx&W\log_2\left(1+\frac{a\rho N_R^2}{b\rho N_R+c
N_R+1}\right),\label{app3}
\end{eqnarray}
where $\mathcal{R}(x)$ denotes the real part of $x$.
$\textbf{h}_{R,D}$ is replaced by
$\sqrt{\rho}\hat{\textbf{h}}_{R,D}+\sqrt{1-\rho}\textbf{e}$ in
(\ref{app1}). (\ref{app2}) follows from the fact that
$\rho\|\hat{\textbf{h}}_{R,D}\|^2$ scales with the order
$\Theta(\rho N_R)$ as $N_R\rightarrow\infty$ while
$2\sqrt{\rho(1-\rho)}\mathcal{R}(\textbf{e}^H\hat{\textbf{h}}_{R,D})
+(1-\rho)\|\textbf{e}\hat{\textbf{h}}_{R,D}^H\|^2/\|\hat{\textbf{h}}_{R,D}\|^2$
scales as the order $\mathcal{O}(1)$, which can be negligible.
(\ref{app3}) holds true because of
$\lim\limits_{N_R\rightarrow\infty}\frac{\|\hat{\textbf{h}}_{R,D}\|^2}{N_R}=1$
and
$\lim\limits_{N_R\rightarrow\infty}\frac{\|\textbf{h}_{S,R}\|^2}{N_R}=1$,
namely channel hardening \cite{ChannelHardening}. Therefore, we get
the Theorem 1.

\section{Proof of Theorem 2}
According to (\ref{eqn5}), given $\varepsilon$, we have
\begin{eqnarray}
\varepsilon&=&P_r\left(C_{SOC}^{AF}>C_{D}^{AF}-W\log_2(1+\gamma_{E}^{AF})\right)\nonumber\\
&=&P_r\left(\gamma_E^{AF}>2^{\left(C_{D}^{AF}-C_{SOC}^{AF}\right)/W}-1\right)\nonumber\\
&=&1-F\left(2^{\left(C_{D}^{AF}-C_{SOC}^{AF}\right)/W}-1\right),\label{app4}
\end{eqnarray}
where $F(x)$ is the cumulative distribution function (cdf) of
$\gamma_E^{AF}$. In order to derive the secrecy outage capacity, the
key is to get the cdf of $\gamma_E^{AF}$. Examining (\ref{eqn11}),
due to channel hardening, we have
\begin{eqnarray}
\gamma_E^{AF}=\frac{dN_R\left|\textbf{h}_{R,E}^H\frac{\hat{\textbf{h}}_{R,D}}{\|\hat{\textbf{h}}_{R,D}\|}\right|^2}
{e\left|\textbf{h}_{R,E}^H\frac{\hat{\textbf{h}}_{R,D}}{\|\hat{\textbf{h}}_{R,D}\|}\right|^2+cN_R+1}.\label{app5}
\end{eqnarray}
Since $\hat{\textbf{h}}_{R,D}/\|\hat{\textbf{h}}_{R,D}\|$ is an
isotropic unit vector and independent of $\textbf{h}_{R,E}$,
$\left|\textbf{h}_{R,E}^H\hat{\textbf{h}}_{R,D}/\|\hat{\textbf{h}}_{R,D}\|\right|^2$
is $\chi^2$ distributed with 2 degrees of freedom. Let
$y\sim\chi_2^2$, we can derive the cdf of $\gamma_E^{AF}$ as
\begin{eqnarray}
F(x)&=&P_r\left(\frac{dN_Ry}{ey+cN_R+1}\leq x\right).\label{app6}
\end{eqnarray}
If $x<dN_R/e$, then we have
\begin{eqnarray}
F(x)&=&P_r\left(y\leq\frac{(cN_R+1)x}{dN_R-ex}\right)\nonumber\\
&=&1-\exp\left(-\frac{(cN_R+1)x}{dN_R-ex}\right).\label{app7}
\end{eqnarray}
Since $x\geq N_R/e$ is impossible when
$x=2^{\left(C_{D}^{AF}-C_{SOC}^{AF}\right)/W}-1$, we have
\begin{equation}
\varepsilon=\exp\left(\frac{(cN_R+1)\left(2^{\left(C_{D}^{AF}-C_{SOC}^{AF}\right)/W}-1\right)}
{dN_R-e\left(2^{\left(C_{D}^{AF}-C_{SOC}^{AF}\right)/W}-1\right)}\right).\label{app8}
\end{equation}
Hence, we get the Theorem 2.

\section{Proof of Theorem 3}
Similarly, according to the definition of secrecy outage capacity in
(\ref{eqn5}), we have
\begin{eqnarray}
\varepsilon&=&P_r(C_{SOC}^{DF}>C_D^{DF}-C_E^{DF})\nonumber\\
&=&P_r\left(\left|\textbf{h}_{R,E}^H\frac{\hat{\textbf{h}}_{R,E}}{\|\hat{\textbf{h}}_{R,D}\|}\right|^2>\frac{2^{(C_D^{DF}-C_{SOC}^{DF})/W}-1}{P_R\alpha_{R,E}}\right)\nonumber\\
&=&\exp\left(-\frac{2^{(C_D^{DF}-C_{SOC}^{DF})/W}-1}{P_R\alpha_{R,E}}\right),\label{app9}
\end{eqnarray}
where (\ref{app9}) follows the fact that
$\left|\textbf{h}_{R,E}^H\frac{\hat{\textbf{h}}_{R,E}}{\|\hat{\textbf{h}}_{R,D}\|}\right|^2$
is $\chi^2$ distributed with 2 degrees of freedom as analyzed
earlier. Based on (\ref{app9}), it is easy to get the Theorem 3.

\end{appendices}

\end{document}